\documentstyle[aps,graphicx,multicol]{revtex}
\begin{document}
\def\btt#1{{\tt$\backslash$#1}}
\draft
\title{{\bf Graphical Representation of 
CP Violation Effects in Neutrinoless Double Beta Decay}}
\author{K. MATSUDA, N. TAKEDA and T. FUKUYAMA}
\address{Department of Physics, 
        Ritsumeikan University, Kusatsu, 
        Shiga 525-8577, Japan}
\author{H. NISHIURA}
\address{Department of General Education, 
        Junior College of Osaka Institute of Technology, 
        Asahi-ku, Osaka 535-8585, Japan}

\date{July 19, 2000}
\maketitle



\begin{abstract}
We illustrate the graphical method that gives the constraints 
on the parameters appearing in the neutrino oscillation experiments 
and the neutrinoless double beta decay.  This method is applicable 
in three and four generations.  Though this method is valid for more 
general case, we examine explicitly the cases in which the $CP$ violating factors 
take $\pm 1$ or $\pm i$  in the neutrinoless double beta decay for illustrative clearance. 
We also discuss some mass matrix models which lead to the above $CP$ violating factors.
\end{abstract}
\pacs{
PACS number(s): 14.60.Pq, 11.30.Er, 23.40.Bw}

\begin{multicols}{2}
From the recent neutrino oscillation experiments\cite{skamioka} 
it becomes affirmative that neutrinos have masses. 
The present and near future experiments enter into the stage of 
precision test for masses and lepton mixing angles. 
In a series of papers we have discussed the relations among
the parameters which appear in the neutrino oscillation experiments and 
in the neutrinoless double beta decay ($(\beta\beta)_{0\nu}$) experiments. 
The most recent experimental upper bound for the averaged mass 
$\langle m_{\nu} \rangle_{e e}$ 
for Majorana neutrinos from $(\beta\beta)_{0\nu}$
is given by $\langle m_{\nu} \rangle_{e e}<0.2 eV$ \cite{baudis} . 
The next generation experiment GENIUS\cite{genius} is anticipated to reach 
a considerably more stringent limit 
$\langle m_{\nu} \rangle_{e e}<0.01-0.001 eV $.
In these situations the investigation of the $CP$ 
violation effects in the lepton sector has 
become more and more important. 
In the previous paper we proposed \cite{matsu} the graphical method 
for obtaining the constraints on $CP$ violation phases from 
$(\beta\beta)_{0\nu}$ and neutrino oscillation experiments. 
This method enables us to grasp the geometrical relations among 
the parameters of masses, mixing angles, $CP$ phases etc. 
and obtain the constraints on them more easily 
than the analytical calculations
\cite{fuku}\cite{fuku2}\cite{nishi}. 
In this letter we apply this method 
to obtain the constraints on $\langle m_{\nu} \rangle_{e e}$  
for the given $CP$ phases.  For illustrative purpose 
we consider the specific values of $CP$ phases which are partly supported by
theoretical models \cite{takasugi}. We also discuss some matrix models which lead to the above $CP$ phases.
\par
The averaged mass $\langle m_{\nu} \rangle_{e e}$ obtained from $(\beta\beta)_{0\nu}$ is given \cite{doi}  
by the absolute values of averaged complex 
masses for Majorana neutrinos as 
\begin{eqnarray}
\langle m_{\nu} \rangle_{e e}  & =& |M_{ee}|.\label{eq761}
\end{eqnarray}
Here the averaged complex mass $M_{ee}$ is defined by
\begin{eqnarray}
M_{ee} 
& \equiv& 
\sum _{j=1}^{3}U_{ej}^2m_j \\
& \equiv&  |U_{e1}|^2m_1+|U_{e2}|^2e^{2i\beta }m_2+|U_{e3}|^2e^{2i(\rho-\phi)}m_3 ,
\end{eqnarray}
after suitable phase convention. 
The $CP$ violating effects are invoked in $\beta$ , $\rho$ and $\phi$ appearing in $U$ . 
Here $U_{a j}$ is the Maki-Nakagawa-Sakata (MNS) left-handed lepton 
mixing 
matrix which combines the weak eigenstate neutrino ($a=e,\mu$ and $\tau$) 
to the mass eigenstate neutrino with mass $m_j$ ($j$=1,2 and 3).  
The $U$ takes the following form in the standard representation 
\cite{fuku}:
\end{multicols}
\hspace{-0.5cm}
\rule{8.7cm}{0.1mm}\rule{0.1mm}{2mm}
\begin{equation}
U=
\left(
\begin{array}{ccc}
c_1c_3&s_1c_3e^{i\beta}&s_3e^{i(\rho-\phi )}\\
(-s_1c_2-c_1s_2s_3e^{i\phi})e^{-i\beta}&
c_1c_2-s_1s_2s_3e^{i\phi}&s_2c_3e^{i(\rho-\beta )}\\
(s_1s_2-c_1c_2s_3e^{i\phi})e^{-i\rho}&
(-c_1s_2-s_1c_2s_3e^{i\phi})e^{-i(\rho-\beta )}&c_2c_3\\
\end{array}
\right).\label{eq772}
\end{equation}
\hspace{9.2cm}
\rule[-2mm]{0.1mm}{2mm}\rule{8.7cm}{0.1mm}
\begin{multicols}{2}
Here $c_j=\cos\theta_j$, $s_j=\sin\theta_j$ 
($\theta_1=\theta_{12},~\theta_2=\theta_{23},~\theta_3=\theta_{31}$). 
Note that three 
$CP$ violating phases, $\beta$ , $\rho$ and $\phi$ appear in $U$ for 
Majorana neutrinos. \cite{bilenky}. 
\par
Defining the relative $CP$ violating factors in $M_{ee}$ by
\begin{equation}
\eta_2 \equiv e^{2i\beta}, \qquad 
\eta_3 \equiv e^{2i(\rho-\phi)} \equiv e^{2i \rho'},
\end{equation}
we have 
\begin{equation}
M_{ee}=|U_{e1}|^2m_1+|U_{e2}|^2\eta_2m_2+|U_{e3}|^2\eta_3m_3
\end{equation}
\par
Let us review briefly the graphical 
representations \cite{matsu} of the complex mass $M_{ee}$. 
We now rewrite the complex mass $M_{ee}$ as
\begin{eqnarray}
M_{ee} 
& =& 
|U_{e1}|^2\widetilde{m_1}+|U_{e2}|^2\widetilde{m_2}+|U_{e3}|^2\widetilde
{m_3}
\label{eq110601}
\end{eqnarray}
Here we have defined the complex masses $\widetilde{m_i} (i=1,2,3)$ by
\begin{eqnarray}
\widetilde{m_1} & \equiv& m_1, \qquad
\widetilde{m_2}  \equiv \eta_2m_2  = e^{2i\beta }m_2, \nonumber\\
\widetilde{m_3} & \equiv& \eta_3m_3  = e^{2i \rho'}m_3,
\end{eqnarray}
\par
The $M_{ee}$ is the "averaged" complex mass of the masses 
$\widetilde{m_i} (i=1,2,3)$ weighted by three mixing elements 
$|U_{e j}|^2  (j=1,2,3)$ with the unitarity constraint $\sum 
_{j=1}^{3}|U_{ej}|^2=1$.
Therefore, the position of $M_{ee}$ in a complex mass plane is 
within the triangle formed by the three vertices 
$\widetilde{m_i} (i=1,2,3)$ if the magnitudes of $|U_{e j}|^2  (j=1,2,3)
$ are unknown (Fig.1). 
This triangle is referred to as the complex-mass triangle\cite{matsu}. 
The three mixing elements $|U_{e j}|^2  (j=1,2,3)$ indicate 
the division ratios for the three portions of each side of the triangle 
which are divided by 
the parallel lines to the side lines of the triangle passing through the 
$M_{ee}.$ (Fig.2).  
The $CP$ violating phases $2\beta$ and $2\rho^\prime$ represent the 
rotation angles of 
$\widetilde{m_2}$ and $\widetilde{m_3}$ around the origin,
respectively. Since $\langle m_{\nu} \rangle_{ee}=|M_{ee}|$, 
the present experimental upper bound on 
$\langle m_{\nu} \rangle_{ee}$ 
(we denote it $\langle m_{\nu} \rangle_{max}$.) 
indicates the maximum distance of the point $M_{ee}$ from 
the origin and forms the circle in the complex plane. 
\par 
So far we have discussed in the scheme of the three flavour mixings. The sterile neutrino is not ruled out yet \cite{lisi}.  We can easily incorporate the sterile neutrino in our formulation.
$M_{ee}$ is written in this case as
\begin{eqnarray}
M_{ee}& =& \sum_{j=1}^4|U_{ej}|^2\widetilde{m_j}
 = |U_{e1}|^2\widetilde{m_1}+|U_{e2}|^2\widetilde{m_2} \nonumber\\
& &+(|U_{e3}|^2+|U_{e4}|^2)\frac{|U_{e3}|^2\widetilde{m_3}+|U_{e4}|^2\widetilde{m_4}}{|U_{e3}|^2+|U_{e4}|^2} \nonumber\\
&\equiv& |U_{e1}|^2\widetilde{m_1}+|U_{e2}|^2\widetilde{m_2}+|U_{e34}|^2\widetilde{m_{34}}
\label{tetragon}
\end{eqnarray}
and Fig.2 is modified as Fig.3.
\begin{figure}[htbp]
\includegraphics{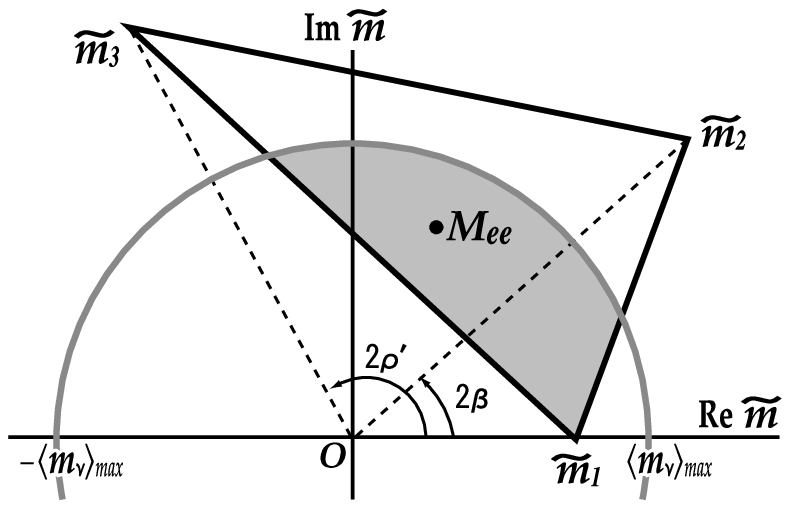}\\
\parbox{8.4cm}{\quad \small FIG.1 \
Graphical representations of the complex mass \(M_{ee}\) 
and $CP$ violating phases.
The position of $M_{ee}$ 
is within the triangle formed by the three points $\widetilde{m_i} (i=1,
2,3)$ which are defined in Eq. (8).
The allowed position of \(M_{ee}\) is in the intersection (shaded area) 
of the inside of this triangle and the inside of the circle of radius 
\(\langle m_\nu \rangle_{max}\) around the origin.}
\label{fig1}
\end{figure}
\begin{figure}[htbp]
\includegraphics{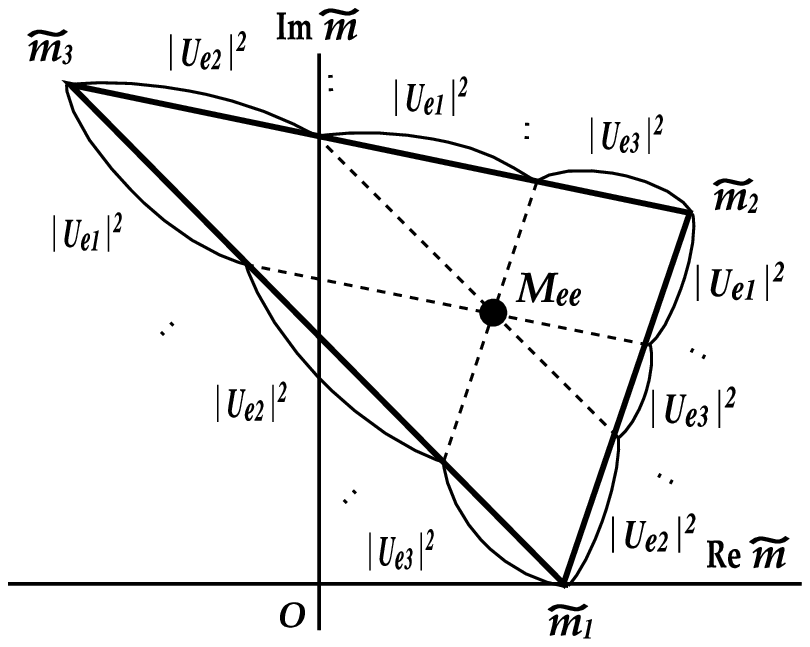}\\
\parbox{8.4cm}{\quad \small FIG.2 \
The relations between the position $M_{ee}$ and $U_{ei} (i=1,2,
3)$ components of MNS mixing matrix.
\vspace{0.3cm} }
\label{fig2}
%
\includegraphics{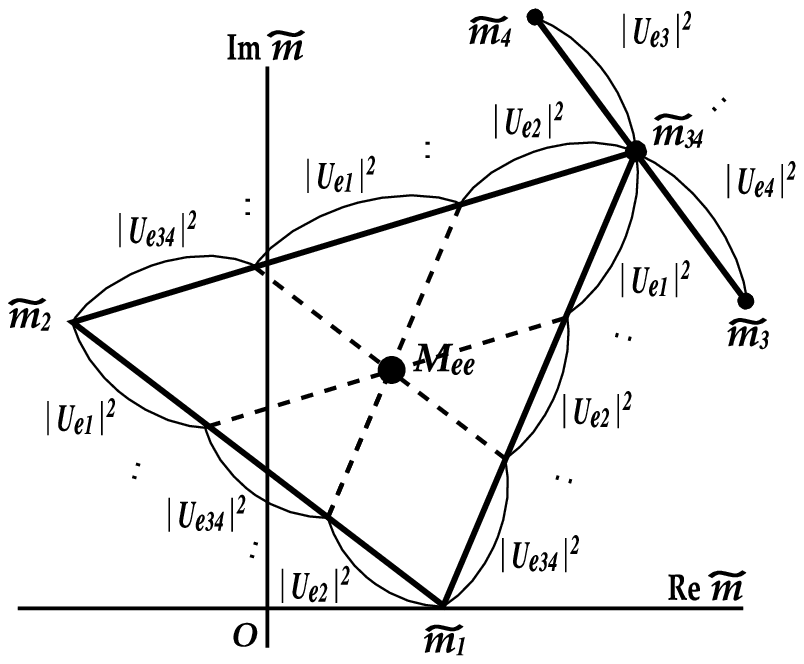}
\parbox{8.4cm}{\quad \small FIG.3 \
Complex-mass "triangle" for four generations.
Complex-mass tetragon formed by the four vertices, $\widetilde{m_j}~~(j=1-4)$, is reduced to "triangle" by (9).
\vspace{0.3cm}}
\label{fig3}
%
\includegraphics{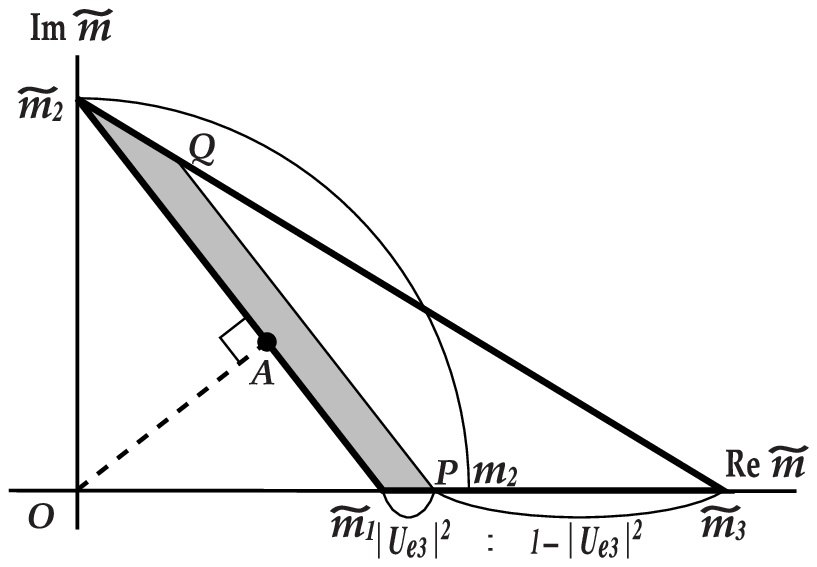}
\parbox{8.4cm}{\quad \small FIG.4 \
Complex-mass triangle supplemented with $|U_{e3}|^2<0.026$ for $\eta_2=i$ and $\eta_3=1$. The shaded region is allowed.
The minimum of $|M_{ee}|$ is $\overline{OA}$. The maximum of $|M_{ee}|$ changes depending on whether $|U_{e3}|_{max}^2$ is larger than $(m_2-m_1)/(m_3-m_1)$ or not.}
\label{fig4}
\end{figure}
\end{multicols}
\newpage
\begin{table}[htbp]
\begin{tabular}{c||ccc}
  {}   & 
  {\(\eta_2=1\)}  & 
  {\(\eta_2=-1\)} & 
  {\(\eta_2=\pm i\)}  \\
 \hline \hline
  \(\eta_3=1\)          & 
     \(m_1 \le \langle m_\nu \rangle_{ee}\) \(\le A_2 \)& 
     \(0 \le \langle m_\nu \rangle_{ee}\) 
      \(\le \mbox{max} 
      \Bigl( 
        \begin{array}{c} 
         m_2 \\ A_1
        \end{array}
      \Bigr)\) & 
     \(B_{12}\le \langle m_\nu \rangle_{ee}\)
     \(\le \mbox{max}
      \Bigl(
       \begin{array}{c} 
        m_2 \\ A_1
       \end{array}
      \Bigr)\)  \\ \hline
  \(\eta_3=-1\) &
     \(\mbox{max}
      \Bigl(
       \begin{array}{c} 
        0 \\ C
       \end{array}
      \Bigr)
      \le \langle m_\nu \rangle_{ee}
      \le \mbox{max}
      \Bigl(
       \begin{array}{c} 
        m_2 \\ -C
       \end{array}
      \Bigr)\) & 
     \( 0 \le \langle m_\nu \rangle_{ee} \le A_2 \) & 
     \(\mbox{max}
      \Bigl(
       \begin{array}{c} 
        0 \\ D_{21}
       \end{array}
      \Bigr)
      \le \langle m_\nu \rangle_{ee} 
      \le \mbox{max}
      \Bigl(
       \begin{array}{c} 
        m_2 \\ E_{23}
       \end{array}
      \Bigr)\)  \\ \hline
   \(\eta_3= \pm i\) & 
     \parbox[t]{5cm}{
      (i) \(|U_{e3}|_{max}^2<m_1^2/(m_1^2+m_3^2)\), \\
      \hspace{0.5cm}
        \(E_{13}\le \langle m_\nu \rangle_{ee} \le
      \mbox{max}
      \Bigl(
       \begin{array}{c} 
        m_2 \\ E_{23}
       \end{array}
      \Bigr)\)\\
       (ii) \(|U_{e3}|_{max}^2\geq m_1^2/(m_1^2+m_3^2)\),\\
       \hspace{0.5cm}
       \(B_{13}\le \langle m_\nu \rangle_{ee} \le
       \mbox{max}
      \Bigl(
       \begin{array}{c} 
        m_2 \\ E_{23}
       \end{array}
      \Bigr)\)
     } & 
     \(0 \le \langle m_\nu \rangle_{ee} \le 
     \mbox{max}
      \Bigl(
       \begin{array}{c} 
        m_2 \\ E_{23}
       \end{array}
      \Bigr)\) & 
     \(B_{12} \le \langle m_\nu \rangle_{ee} \le A_2\) 
      \\ \hline
   \(\eta_3=\mp i\)  & \rule{3cm}{0.1mm} & \rule{3cm}{0.1mm} & 
     \(\mbox{max}
           \Bigl(
       \begin{array}{c} 
        0 \\ D_{12}
       \end{array}
      \Bigr)\le \langle m_\nu \rangle_{ee} 
     \le \mbox{max}
      \Bigl(
       \begin{array}{c} 
        m_2 \\ E_{23}
       \end{array}
      \Bigr)\) 
\end{tabular}
\parbox{17.8cm}{\small \quad TABLE I. \
Given the \(\eta_i=\pm 1\) or \(\pm i\), 
the constraints on the averaged mass \(\langle m_{\nu}\rangle_{ee}\) 
are given. 
Notations are as follows. \(|U_{e3}|_{max}^2\) \(=0.026\), 
\(A_i \equiv\) \(m_i+|U_{e3}|_{max}^2(m_3-m_i)\), 
\(B_{ij} \equiv m_im_j/\sqrt{m_i^2+m_j^2},\)
\(C\equiv\) \(m_1-|U_{e3}|_{max}^2(m_1+m_3)\)
\(D_{ij} \equiv m_i(m_j-|U_{e3}|_{max}^2(m_j+m_3))/\sqrt{m_i^2+m_j^2}\) and 
\(E_{ij} \equiv \sqrt{(1-|U_{e3}|_{max}^2)^2m_i^2+|U_{e3}|_{max}^4 m_j^2}\). 
The max \((a,b)^T\) indicates the larger value 
between \(a\) and \(b\). The double signs of \(\eta_2\) and \(\eta_3\) are in
the same order. }
\label{table1}
\end{table}
%
%
%
\begin{figure}[htbp]
\includegraphics{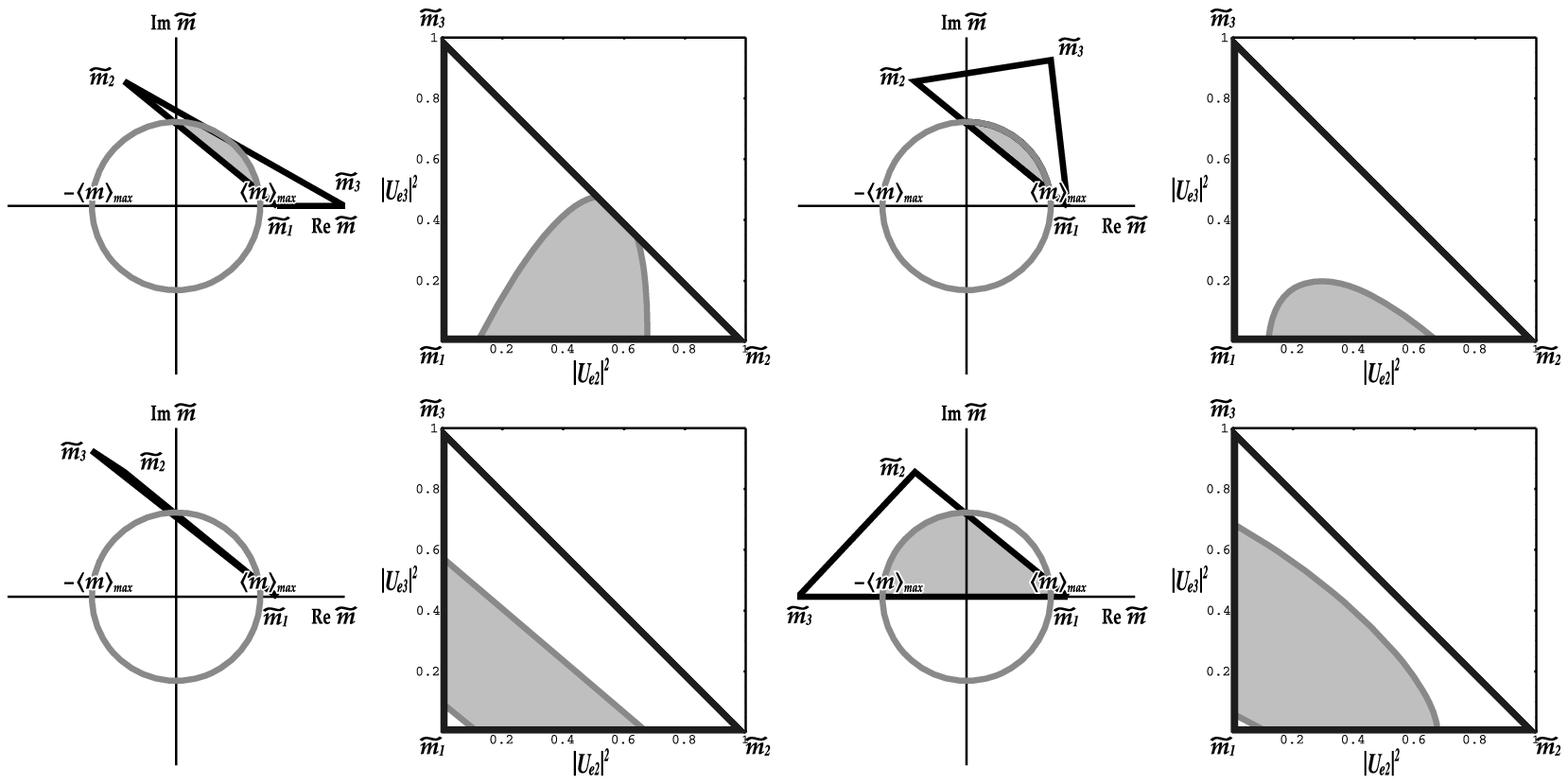}
\parbox{17.8cm}{\small \quad FIG.5 \
The graph shows the allowed region of \(|U_{e2}|^2\) and  \(|U_{e3}|^2\). 
The mixing elements $|U_{ei}|^2$ are expressed as the ratios of the heights 
from vertices $m_i$ separated by $M_{ee}$ (Fig.2). 
The mass triangle (left diagrams) is deformed to the isosceles right triangle (right diagrams) retaining these ratios. 
This figure is depicted setting \(m_1:m_2:m_3:\langle m \rangle_{ee} =6:8:10:5\) and \(2\beta=5\pi/8\) as an example. \(2\rho'\) is changed every \(\pi/3\).
There is no allowed region in upper-right half, 
because \(|U_{e2}|^2+|U_{e3}|^2=1-|U_{e1}|^2 \le 1\) 
from unitarity conditions.
}
\label{fig5}
\end{figure}
\begin{multicols}{2}
\par 
We proceed to discuss the main theme to
give the constraints on $\langle m_{\nu} \rangle_{ee}$ , given $\eta_i=\pm 1$ or $\pm i$.  
We argue here in the three generations, though the arguments can be extended to the four generation case.  
We list up the constraints on $\langle m_{\nu} \rangle_{e e}$  
for the various combinations of given $CP$ violating factors $\eta_i$ and explain how our formulation works. (TABLE I)

We demonstrate the graphical method for $\eta_2=i$ and $\eta_3=1$ case.
In this case Fig.1 becomes Fig.4.
Here we impose the constraint on $U_{e3}$ from the oscillation experiments of CHOOZ and SuperKamiokande \cite{chooz}, $|U_{e3}|^2<0.026$.  The shaded region is allowed.  $\langle m_{\nu} \rangle_{e e}$ is the distance from the origin to the shaded region.  So the minimum and maximum of $\langle m_{\nu} \rangle_{e e}$ is easily estimated from Fig.4.
Obviously, the minimum is $\overline{OA}$.  
The maximum value depends on whether 
the line PQ crosses the horizontal axis at a point larger than $m_2$ or not.
If $\overline{OP} >m_2$, the maximum is $\overline{OP}$.  If $\overline{OP} <
m_2$, the maximum becomes $m_2$.
\par
Of course, our method is applicable for arbitrary phases.  Using Figs.1 and 2,
we can provide the constraints on the mixing elements $|U_{ei}|$ by changing 
the complex-mass triangle to the isosceles right triangle (Fig.5).  
\par 
Next let us discuss a mass matrix model which gives the typical 
$CP$ violating phases demonstrated above.
In order to see the meaning of $\eta=\pm 1$ or $\pm i$ clearly, we first discuss in two generations.
We show that the following democratic type of complex mass matrix,
\begin{eqnarray}
\left(
\begin{array}{cc}
M&M^*\\
M^*&M\\
\end{array}
\right)
\end{eqnarray}
gives the $CP$ violating factor $\eta_2= i$ . 
Here the matrix elements $M^*$ is the complex conjugate of $M$. 
This matrix is diagonalized by a maximal mixing matrix and 
two mass eigen values remain to be independent free parameters as is shown
\begin{eqnarray}
&&\left(
\begin{array}{cc}
\frac{1}{\sqrt{2}}&\frac{1}{\sqrt{2}}\\
-\frac{1}{\sqrt{2}}&\frac{1}{\sqrt{2}}\\
\end{array}
\right)\left(
\begin{array}{cc}
M&M^*\\
M^*&M\\
\end{array}
\right)\left(
\begin{array}{cc}
\frac{1}{\sqrt{2}}&-\frac{1}{\sqrt{2}}\\
\frac{1}{\sqrt{2}}&\frac{1}{\sqrt{2}}\\
\end{array}
\right)\nonumber\\
&&\hspace{1cm}=\left(
\begin{array}{cc}
m_1&0\\
0&im_2\\
\end{array}
\right),
\end{eqnarray}
where
\begin{equation}
m_1\equiv 2\mbox{Re}M, \quad m_2\equiv 2\mbox{Im}M.
\label{mdef}
\end{equation}
\par
This argument is generalized to the three generations as follows.
We consider a model in which the mass matrices for the charged leptons and 
the Majorana neutrinos, $M_{e}$ and
$M_{\mu}$, are  given by
\begin{eqnarray}
M_{e}&=&\mbox{diag}(m_e,m_{\mu},m_{\tau}),\\
M_{\nu}&=&\left(
\begin{array}{ccc}
c_3&0&s_3\\
-s_2s_3&-c_2&s_2c_3\\
-c_2s_3&s_2&c_2c_3\\
\end{array}
\right) \nonumber \\
&&\times
\left(
\begin{array}{ccc}
M&M^*&0\\
M^*&M&0\\
0&0&m_3\\
\end{array}
\right)\left(
\begin{array}{ccc}
c_3&0&s_3\\
-s_2s_3&-c_2&s_2c_3\\
-c_2s_3&s_2&c_2c_3\\
\end{array}
\right)^{T}.\label{eq190}
\end{eqnarray}
Then, the mass matrix $M_{\nu}$ is 
diagonalized by an orthogonal matrix $O_{\nu}$ with 
a pure imaginary eigenvalue for the second generation neutrino as 
\begin{equation}
O_{\nu}^TM_{\nu}O_{\nu}=
\left(
\begin{array}{ccc}
m_1& & \\
 &im_2& \\
 & &m_3\\
\end{array}
\right),\label{eq197}
\end{equation}
where
\begin{equation}
O_{\nu}=
\left(
\begin{array}{ccc}
c_1c_3&s_1c_3&s_3\\
-s_1c_2-c_1s_2s_3&
c_1c_2-s_1s_2s_3&s_2c_3\\
s_1s_2-c_1c_2s_3&
-c_1s_2-s_1c_2s_3&c_2c_3\\
\end{array}
\right).\label{eq200}
\end{equation}
with the maximal mixing for $\theta_1$. 
Namely the assumption Eq.(\ref{eq190}) leads to a pure imaginary 
eigenvalue $im_2$ and a maximal mixing $c_1=s_1=\pi/4$. 
Here $m_1$ and $m_2$ are defined in (\ref{mdef}).
The mass matrix is finally diagonalized with real diagonalized masses as 
\begin{equation}
U_{\nu}^TM_{\nu}U_{\nu}=
\left(
\begin{array}{ccc}
m_1& & \\
 &m_2& \\
 & &m_3\\
\end{array}
\right),\label{eq220}
\end{equation}
where
\begin{equation}
U_{\nu}=O_{\nu}
\left(
\begin{array}{ccc}
1& & \\
 &e^{i\frac{\pi}{4}}& \\
 & &1\\
\end{array}
\right).\label{eq230}
\end{equation}
This mass matrix model corresponds to the case with 
$\eta_2=i$ and $\eta_3=1$ discussed before.\\

The work of K.M. is supported by the JSPS Research Fellowship, No.10421.


%
\end{multicols}

\end{document}